\documentclass[]{aanew}

\begin{document}

\title{The population of massive X-ray binaries\thanks{Based 
on observations collected at the European Southern Observatory,
Chile (ESO 64.H-0059, ESO 66.D-0299) and the South African
Astronomical Observatory}}
\subtitle{I. The Large Magellanic
  Cloud}

\author{I.~Negueruela\inst{1}
\and M.~J.~Coe\inst{2}
}                   
                                                            
\institute{Observatoire de Strasbourg, 11 rue de l'Universit\'{e},
F67000 Strasbourg, France
\and
Physics and Astronomy Department, Southampton University, 
Southampton SO17 1BJ, U.K.
}

\offprints{I. Negueruela,
\email{ignacio@astro.u-strasbg.fr}}

\date{Received    / Accepted     }

\titlerunning{LMC X-ray binaries}
\authorrunning{Negueruela \& Coe}

\abstract{
We present high resolution blue spectroscopy of an almost complete sample of
optical counterparts to massive X-ray binaries in the Large Magellanic
Cloud (LMC) and derive their spectral classification. We find an
spectral type B0II for the optical counterpart to \object{RX
J0532.5$-$6551}, confirming it as the first wind-fed massive X-ray
binary in the LMC. We also confirm the Be nature of the proposed
counterpart to \object{RX J0535.0$-$6700}. The proposed optical
counterpart to  \object{RX J0531.5$-$6518} is a B2V star with signs of
emission in the Balmer lines. In total, we give accurate spectral
types for 14 counterparts. We find that the 
overall observed population of massive X-ray binaries in the LMC has a
distribution not very different from the observed Galactic population
and we discuss different selection effects affecting our knowledge of
this population.
The spectral distribution of the Be/X-ray binary population is also
rather similar to the Galactic one. This distribution implies that
Be/X-ray binaries must have preferentially formed from moderately
massive binaries undergoing semi-conservative evolution. The
observation of several Be/X-ray binaries with large eccentricities
implies then the existence of supernova kicks.
}
\maketitle 


\section{Introduction}

High Mass X-ray Binaries (HMXBs) are X-ray sources composed of an early-type 
massive star and an accreting compact object (generally a neutron star, but 
occasionally a black hole). HMXBs are traditionally divided 
(see Corbet 1986) into Classical or
Supergiant X-ray  binaries (SXBs) in which the compact object 
accretes from  a mass-losing OB supergiant or bright giant and
Be/X-ray binaries (BeXBs),  
in which a neutron star orbits an unevolved OB star surrounded by a dense 
equatorial disc (cf. Liu et al. 2000 for a recent catalogue), though 
the physical reality could be rather more complex (cf. Negueruela \&
Reig 2001). Different population synthesis analyses predict that the vast
majority of HMXBs will be BeXBs, though this is not apparent from the
number of sources detected in the Milky Way (where $\sim 30$\% of
known systems are SXBs), presumably due to different selection effects
that will be detailed later on.

Apart from their intrinsic interest as high-energy radiation sources,
HMXBs offer a window on the late stages of massive binary
evolution. Their properties as a population can be used to extract
valuable information on the different stages of the life of massive
binaries. 

In order to understand the representativity of the known sample of
Galactic HMXBs, it is of fundamental importance to have at least an
approximate idea of the HMXB population content in other Galaxies. The
Magellanic Clouds (MCs) present a unique opportunity to carry out such
a study, since they have a structure and chemical composition which
differs from that of the Milky Way and, at the same time, are close
enough to allow the study of individual sources with modest-sized
ground-based telescopes.

For this reason, we have undertaken an observational campaign in order
to obtain high Signal-to-Noise Ratio (SNR) spectroscopy of the optical
counterparts to MC HMXBs which will allow us to derive accurate
spectral classifications for these stars. Such work is imperative if
we are to gain some understanding of the mass distribution and
evolutionary status of the HMXB populations.

In this first paper, we centre on the HMXB population of the LMC. This
population is relatively small and all the optical counterparts are
reasonably bright, allowing us to obtain a basically complete
sample. In further works, we will study the SMC population and will
compare the characteristics of the MC populations with the Galactic
one.

\section{Sample and Observations}

Spectra in the classification region were obtained using
the ESO 1.52-m telescope at La Silla Observatory, Chile, equipped with 
the Boller \& Chivens spectrograph. On the nights of
30th and 31st October 1999, the spectrograph was fitted with
the \#32 holographic grating and the Loral \#38 camera, which 
gives a nominal resolution of $\sim 0.5 $\AA/pixel. On 1st November 
1999 and 15th September \& 22nd October 2000, we used the \#33 holographic 
grating instead, which gives a resolution of
$\sim 1.0$ \AA/pixel.
Measurements of arc line FWHMs  indicate spectral
resolutions of $\approx 1.4 $\AA\ and $\approx 3.0 $\AA\ 
at $\sim 4500$\AA, respectively.

\begin{table*}[ht]
\caption{Photometric observations of the sources discussed. For
  EXO~053109$-$6609 see text.}
\begin{center}
\begin{tabular}{llcccc}
Object&Date&$B$&$V$&$R$&Error\\
\hline
RX J0532.5$-$6551 &4 Oct 1996     &13.00  &13.09  &13.17  &0.01   \\
LMC X-4         &6 Oct 1996     &14.38  &14.44  &14.86  &0.03   \\
LMC X-3         &20 Jan 2001    &16.85  &16.74  &16.89  &0.05   \\
CAL E           &3 Oct 1996     &14.35  &14.42  &14.44  &0.01   \\
CAL 9           &2 Oct 1996     &14.48  &14.36  &14.23  &0.02   \\
RX J0520.5$-$6932 &4 Oct 1996     &14.42  &14.40  &14.32  &0.03   \\
RX J0544.1$-$7100 &24 Jan 1999    &15.35  &15.25  &15.18  &0.01   \\
RX J0529.8$-$6556 &25 Jan 1999    &14.65  &14.81  &14.96  &0.01   \\
RX J0531.5$-$6518 &21 Jan 1999    &15.86  &16.02  &16.11  &0.02   \\
H0544$-$665       &1 Oct 1996     &15.20  &15.55  &15.26  &0.01   \\
1A 0535$-$66      &20 Jan 2001    &14.91  &14.90  &15.30  &0.04   \\
RX J0535.0$-$6700 &19 Jan 1999    &14.80  &14.87  &14.91  &0.01   \\
\hline
\end{tabular}
\end{center}
\label{tab:phot}
\end{table*}

The configurations selected introduce a magnitude limit (the magnitude
at which a 40-min integration with the \#33 grating 
would not give an adequate SNR for
accurate spectral classification) at $B\la16$. The only HMXB in the LMC
whose optical counterpart is fainter than this limit is LMC X-3. We obtained
a lower resolution spectrum of this source on 10th February 2001 with
the Danish 1.54-m telescope at La Silla Observatory, Chile, equipped with 
the Danish Faint Object Spectrograph and Camera (DFOSC) and grism \#3.

All the data have been reduced with the {\em Starlink} packages {\sc
ccdpack} \cite{draper} and {\sc figaro} \cite{shortridge} and analyzed
using {\sc figaro} and {\sc dipso} \cite{howarth}.
Photometry for some of the objects has been obtained as part of the
Southampton/SAAO HMXB campaign. It has been taken during several
runs using the SAAO 1.0-m telescope. The 
photometric data are listed in Table~\ref{tab:phot}.

Our sample is complete, in the sense that we present accurate
spectral classifications for the 14 LMC MXBs with secure optical
identifications listed in the catalogue of Liu et al. (2000). Recent
deep surveys of {\em ROSAT} sources in LMC fields have failed to
detect many new HMXB candidates, in spite of intensive work (Haberl \&
Pietsch 1999a; Sasaki et al. 2000). Three
other massive stars have been suggested as counterparts to possible
MXBs, but they have not been included in the sample. Among this, the
identification of \object{RX J0516.0$-$6916} with a $V=15.0$ B1V star was
suggested by Cowley et al. (1997), but considered uncertain because
the star did not display any characteristics of Be behaviour at the
time of the observations. Sasaki et al. (2000) suggest the
identification of \object{RX J0541.4$-$6936} with the LMC supergiant
\object{Sk $-69\degr$ 271} (whose magnitude is listed as $B=11.6$, and not 
$B=18.8$ as given by Sasaki et al.) because of positional coincidence.
Finally, \object{RX J0532.4$-$6535} was suggested by Haberl \& Pietsch
(1999) as a Be/X-ray binary, because of its coincidence with a
variable star in the catalogue of Reid et al. (1988), namely {\object
  GRV\,0532$-$6536}. Two other
{\em ROSAT} PSPC sources which show positional coincidence with stars
in this catalogue turn out to be Be/X-ray binaries (see later the
discussion on \object{RX J0535.0$-$6700}), even though objects
in this catalogue were originally believed to be Mira
variables. However, the coulour given by Reid et al. (1988) for 
{\object GRV\,0532$-$6536} $(V-I)=2.02$ strongly suggests that it is
indeed a red star and not an LMC Be star.

\section{Spectral classification}
\label{sec:spec}

Due to the smaller metal content of the LMC with respect to the Milky
Way, spectral classification of LMC sources cannot be achieved with a
straightforward comparison to Galactic MK standards. Though some
effort has been done in establishing a classification scheme for the
brightest LMC supergiants (e.g., Fitzpatrick 1991), this work has not been
extended to lower luminosities.

The metallic lines in LMC supergiants are on average 30\% weaker than
in Galactic objects of the same spectral type and luminosity class,
but there is some dispersion \cite{fit91}. Such a decrease in
intensity may mean
that what appears as a weak line in the spectrum of a bright Galactic
MK standard is lost in the noise in the spectrum of an LMC source. 
This complicates
specifically the luminosity classification. Since the spectral
classification (and specially the luminosity classification) of Be
stars is already difficult for Galactic objects due to the presence of
emission components (see Steele et al. 1999 for a general discussion),
great care has to be taken when obtaining spectral types for the
counterparts to MXBs in the LMC.

\begin{figure*}
\begin{picture}(500,240)
\put(0,0){\includegraphics{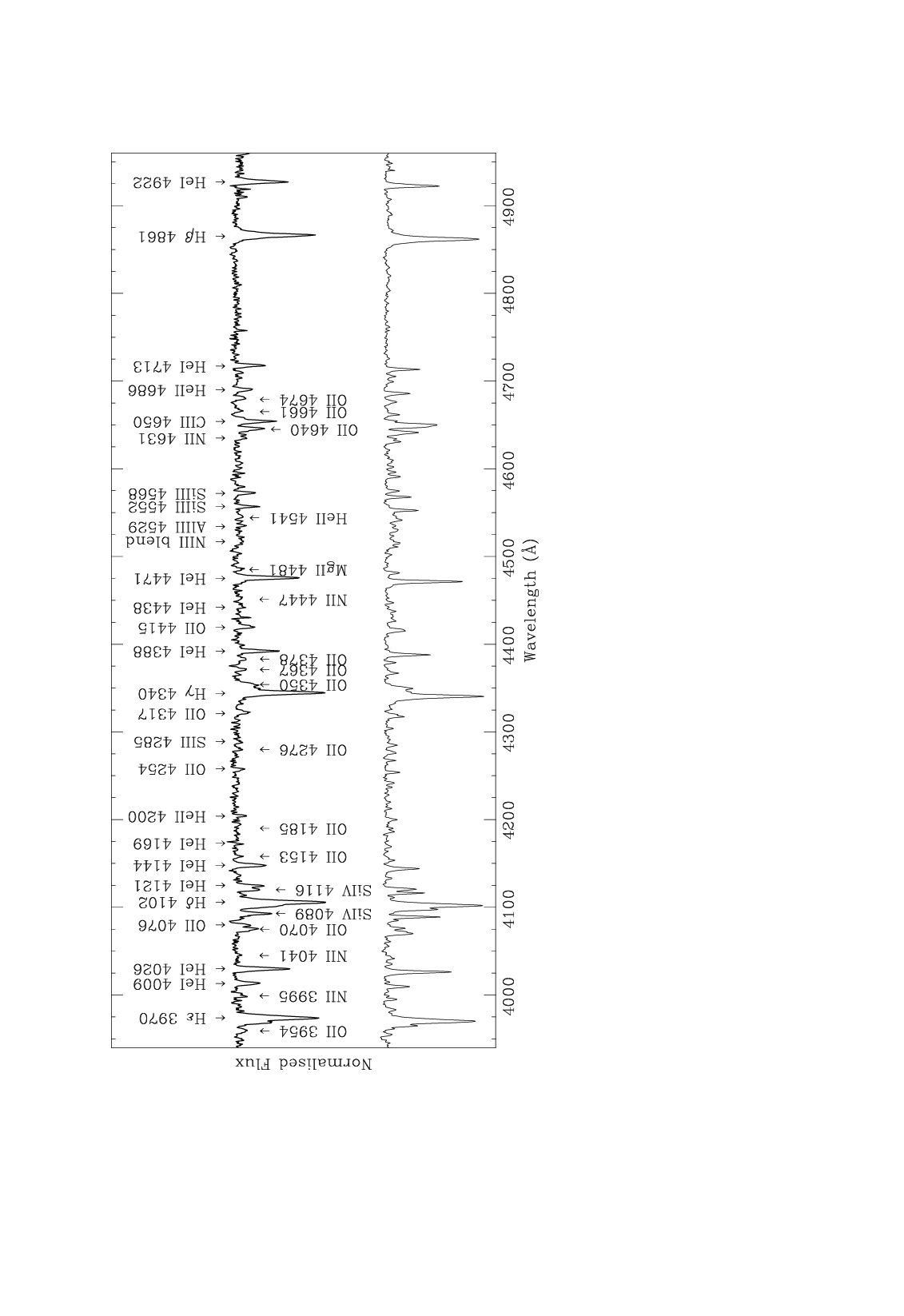}}
\end{picture}
\caption{Blue spectrum of \object{Sk $-65\degr$ 66} (top), compared to that of
  the B0III MK standard \object{HD 48434} 
  observed with the same instrumentation. Features detected on
  the spectrum of  \object{HD 48434}  have been marked on the spectrum
  of \object{Sk $-65\degr$ 66} in order to facilitate comparison.
  Note the weaker
  \ion{He}{ii}~$\lambda$4686\AA\ in the spectrum of  \object{Sk $-65\degr$
  66} and the general shift in wavelength of all the lines due to the
  LMC systemic radial velocity.}
\label{fig:superspec}
\end{figure*}

For this reason we have adopted the criterion of assuming that all Be
stars are main-sequence objects unless the intrinsic magnitudes
derived from photometry of the objects suggest a higher luminosity
class. In what follows, we
will adopt the value of the distance modulus derived by Udalski (2000)
from the results of the OGLE-II microlensing experiment, i.e.,
$(M-m)_{0}=18.24$. Since this value is considerably smaller than
values adopted before, the effect of this choice will be discussed in Section~\ref{sec:dis}.

One main source of difficulty stems from the fact that the
circumstellar disk around a Be star can contribute to its
luminosity. In principle, since the disks have relatively
low temperatures compared to the central stars (typically $T_{{\rm
    disk}} \approx \frac{1}{2} T_{{\rm eff}}$), the disk contributes
significantly to the continuum emission (and hence the total
luminosity) only at relatively long wavelengths (near- and
mid-IR), giving rise to what is generally known as {\em infrared
  excess} (e.g., Slettebak 1988). However, variability by up to $\sim
0.5\:{\rm mag}$ in the $V$ and $R$ bands is not infrequent in isolated
(i.e., without a binary companion) Be stars and variability in the $B$
band can be substantial, even if it is not expected from modelling
\cite{dac}.

\subsection{\object{RX J0532.5$-$6551} = \object{Sk $-$65$\degr$ 66}}

\object{RX J0532.5$-$6551} is a very variable (up to a factor of 66; Sasaki et
al. 2000), apparently persistent, not very bright 
($L_{{\rm x}}\sim 1\times10^{35}\:\mathrm{erg}\,\mathrm{s}^{-1}$)
X-ray source, associated with the bright star \object{Sk $-65\degr$ 66}
(Haberl et al. 1995a). All these characteristics make it the only LMC
X-ray source proposed to be an analogue to the Galactic wind-fed
SXBs. In Fig.~\ref{fig:superspec}, we present the classification
spectrum of \object{Sk $-65\degr$ 66}, together with that of the B0III
MK standard \object{HD 48434}.

The spectrum of \object{Sk $-65\degr$ 66} displays weak \ion{He}{ii} lines,
indicating a spectral type close to B0. The richness of the metallic
spectrum would prevent a main-sequence classification even for a
Galactic object. In Fig.~\ref{fig:superspec}, we can see that the
strength of the oxygen spectrum in \object{Sk $-65\degr$ 66} is
comparable to the B0III Galactic standard, while the \ion{Si}{iii} and
\ion{Si}{iv} lines are slightly weaker. The strength of the \ion{H}{i}
and He lines (which are not affected by metallicity) is similar to
that of the B0III standard, though always slightly weaker. The
\ion{He}{ii}~$\lambda$4686\AA\ line, which is the main luminosity
indicator, is clearly weaker in \object{Sk $-65\degr$ 66}, suggesting a
slightly higher luminosity. Therefore the spectrum indicates a
spectral type of B0II for \object{Sk $-65\degr$ 66}.

For this spectral type, Wegner (1995) gives an intrinsic colour
$(B-V)_{0}=-0.22$. Our photometric data, $V=13.09$,
$(B-V)=-0.09\pm0.02$ then imply a reddening $E(B-V)=0.13$, consistent
with the average for LMC sources. Assuming standard extinction
$A_{V}=3.1E(B-V)$, this results in an intrinsic luminosity
$M_{V}=-5.6$, which is too high for a B0III star (Vacca et
al. 1996), but  compatible with a B0II or B0Ib spectral type. We
conclude then that \object{Sk $-65\degr$ 66} has a spectral type B0II,
confirming it as the first wind-fed SXB in the LMC.

\subsection{\object{LMC X-4}}

 \begin{figure*}
\begin{picture}(500,240)
\put(0,0){\includegraphics{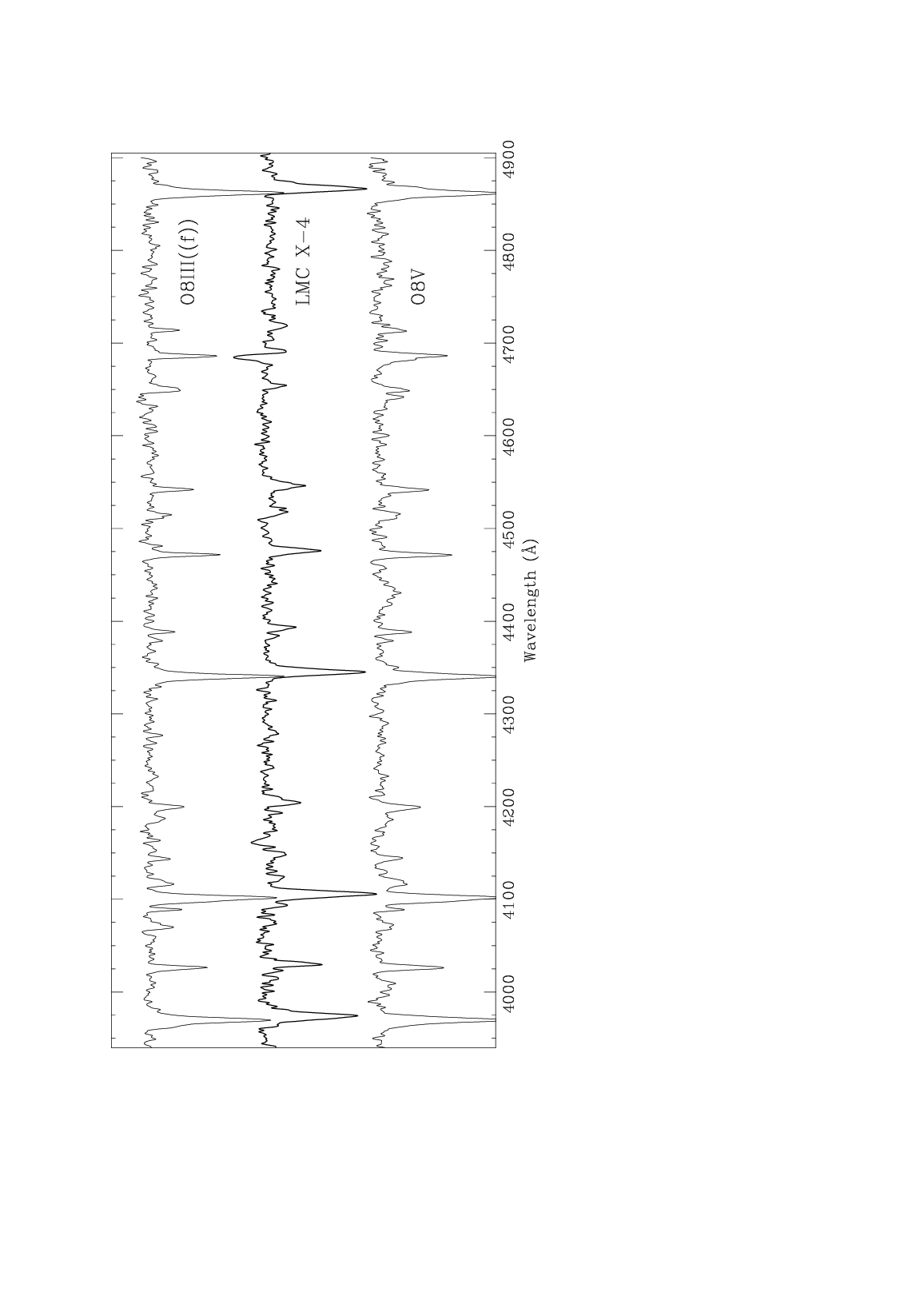}}
\end{picture}
\caption{Blue spectrum of the optical counterpart to \object{LMC X-4},
  (middle) compared to MK standard stars for spectral types O8III((f))
  (HD36861) and O8V (HD48279). The apparent emission feature around
  $\approx 4160$ in the spectrum of LMC X-4 is due to a
  cosmic ray hit. Note the photospheric component of
  \ion{He}{ii}~$\lambda$4686\AA, clearly visible in spite of the
  strong emission from the vicinity of the compact object. The
  spectra, as in all the following figures, have been smoothed with a
  Gaussian filter for display.}
\label{fig:lmcx4}
\end{figure*}

This very bright (up to $\sim 5\times10^{38}\:{\rm erg}\,{\rm
s}^{-1}$), variable and persistent X-ray source is a
13.5-s X-ray pulsar (Kelley et al. 1983). The system has been
extensively studied and it is understood in terms of a neutron star in
a 1.4-d orbit around a massive O-type star close to filling its Roche
lobe (Chevalier \& Ilovaisky 1977). A strong 30-d modulation in the
optical and X-ray lightcurves is identified in terms of a precessing
accretion disc (e.g., Heemskerk \& van Paradijs 1989).

Chevalier \& Ilovaisky (1977; see also Ilovaisky et al. 1984) observed
photometric variability with an amplitude of $\sim 0.2$ mag in the
$B$-band and some changes in the width and strength of lines in the
spectrum in phase with the 1.4-d orbital period. The spectral class,
however, was approximately constant in spite of the expected X-ray
heating of the O-star surface. Their average spectrum was that of an
O8III--V star with weak variable \ion{He}{ii}~$\lambda$4686\AA\ emission
superimposed on the stellar spectrum. The radial velocity of the
emission line varied strongly, indicating that it is produced close to
the compact object. Further spectroscopic observations by Hutchings et
al. (1978) found very weak or absent \ion{He}{ii}~$\lambda$4686\AA\
emission, but suggested changes in the spectral type with the orbital
period. 

Our spectrum of \object{LMC X-4} is displayed in Fig.~\ref{fig:lmcx4} together
with MK standard stars for spectral types O8III((f)) (HD36861)
and O8V (HD48279), taken from the digital atlas of Walborn \&
Fitzpatrick (1990). Even though there is relatively strong
\ion{He}{ii}~$\lambda$4686\AA\ emission (much stronger than in any of the
spectra of Hutchings et al. 1978), the underlying photospheric feature
is clearly visible. As indicated by Chevalier \& Ilovaisky (1977), the
spectrum seems intermediate between O8V and O8III, with the strength
of the He lines closer to the giant. The ratio
\ion{Si}{iv}~$\lambda$4089\AA/\ion{He}{i}~$\lambda$4143\AA, which is
the main luminosity indicator for Conti \& Alschuler (1971) also seems
intermediate, but closer to the giant. The fill-in and possibly
marginal emission in \ion{N}{iii}~$\lambda$4640\AA\ should confirm that the
star is relatively evolved, as its position in the HR diagram
indicates (cf. Savonije 1980), though this emission could originate close
to the compact object.

Therefore a spectral type O8III seems adequate.
Ilovaisky et al. (1984) determined the average colour of \object{LMC
X-4} to be $(B-V)= -0.23$, consistent with the $E(B-V)=0.05$ derived
from ultraviolet observations. Then $B=13.96$
(average of the X-ray off states) implies $M_{V}=-4.2$. We have then
to conclude that the source is underluminous for its spectral type
-- even for O8V, Vacca et al. (1996) give $M_{V}=-4.7$. One
possible explanation may be that the observed spectral type is due to
the deformation and heating of a star of considerably lower mass than
corresponds to this spectral type by the compact companion.

\subsection{\object{LMC X-1}}

The optical counterpart for this X-ray source was finally
identified with an O-type star (\object{star \#32}) by Cowley et
al. (1995), based on {\em ROSAT} observations. Previous positions did
not allow to choose between this star and a nearby  B
supergiant. Hutchings et al. (1983) showed that the O star was a
spectroscopic binary with a likely 4.2-d period, suggesting a massive
companion. This star is surrounded by a bright emission nebula, likely
to be partly ionized by the X-ray flux \cite{paa86}. The lack of
pulsations and the soft X-ray spectrum make \object{LMC X-1} a clear
black-hole candidate. The X-ray source is persistent and the
luminosity is high ($\approx 2\times10^{38}\:{\rm erg}\,{\rm s}^{-1}$).

\begin{figure*}
\begin{picture}(500,240)
\put(0,0){\includegraphics{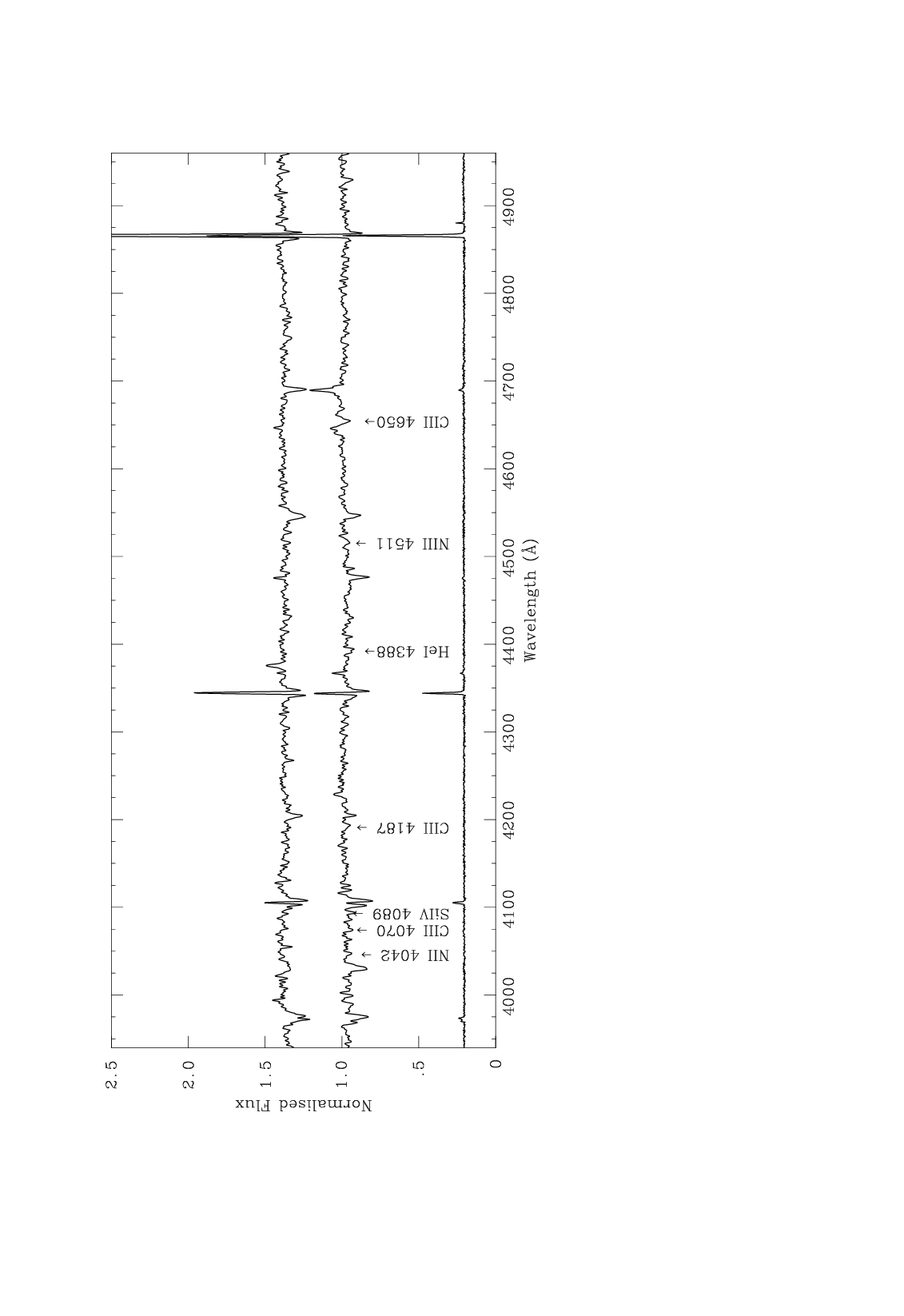}}
\end{picture}
\caption{Blue spectrum of the optical counterpart to \object{LMC X-1}
  (middle). The top spectrum is that of a second O-type star lying on
  the slit.  The scale is set so that the photospheric features are clearly
  visible. The bottom spectrum is a sky spectrum taken very close to
  the position of \object{LMC X-1}, showing the very bright emission
  lines produced in the nebula surrounding the star (arbitrarily scaled).
  See Fig.~\ref{fig:lmcx4} for standards of similar spectral type.}
\label{fig:lmcx1}
\end{figure*}

The spectrum of \object{star \#32} is displayed in
Fig.~\ref{fig:lmcx1}, together with that of a second O-type star
immersed in the nebulosity. The spectrum is dominated by the very bright
emission lines from the surrounding nebula and has been scaled so as
to display the photospheric features clearly. These are very weak in
both stars, but even more so in the spectrum of \object{star
  \#32}. The absolute weakness of the \ion{He}{ii} lines (apart from
\ion{He}{ii}~$\lambda$4686\AA\ which has a strong nebular component)
is unlikely to be due to in-filling by nebular emission (since the sky
spectrum does not show these lines in emission -- see
Fig.~\ref{fig:lmcx1}) and may suggest that
there is some contribution of the accretion disk around the compact
object to the continuum.

Though the ratio \ion{He}{ii}~$\lambda$4541\AA $\simeq$
\ion{He}{i}~$\lambda$4471\AA\ implies a spectral type close to O7, the
strength of the \ion{He}{i} line is probably affected by nebular emission
(see the spectrum of the second O-type star where this line is seen in
emission). The possible presence of
\ion{C}{iii}~$\lambda$4187\AA\ and relatively strong
\ion{Mg}{ii}~$\lambda$4481\AA\ favours a spectral type O8III (see
Fig.~\ref{fig:lmcx4}). Assuming that the reddening to the star
has a component $E(B-V)=0.05$ due to Galactic foreground absorption
and $E(B-V)=0.32$ due to the surrounding nebulosity \cite{bp85}, an
absolute magnitude $M_{V}=-4.9$ is derived. This is slightly too
bright for a main-sequence O8 star and slightly too faint for a giant,
in good agreement with the spectral classification (given the
uncertainty in the reddening law that should be used for the nebular
component). Our photometric
data from 1996 are also compatible with these values.

\subsection{\object{LMC X-3}}

\object{LMC X-3} is a very bright X-ray source ($L_{\rm x}\approx
10^{38}\:{\rm erg}\,{\rm s}^{-1}$). An orbital solution for
\object{LMC X-3} was first proposed by Cowley et al. (1983),
based on spectroscopy of the optical counterpart, for which they suggested
a spectral type B3V. They derived a 1.70-d period and a 
mass for the compact object $M_{\rm x} > 7.0\: M_{\sun}$ and likely 
$9\:M_{\sun}$, clearly indicating that it is a massive black hole. From
photometric observations, van der Klis et al. (1985) were able to refine
the orbital period to $P=1.70479\:{\rm d}$ and constrain the black
hole mass to $9\:M_{\sun} < M_{\rm x} < 13\:M_{\sun}$. Such values imply
 $M_{\rm x} > M_{*}$, a situation in which stable 
mass transfer via Roche-lobe accretion can take place. 
However, Soria et al. (2001) have argued that the optical 
star must have a spectral type B5IV in order to fill its Roche lobe. 

The optical counterpart to \object{LMC X-3} is very variable, changing between
$V=16.5$ and $V=17.3$, indicating that optical emission from the
vicinity of the compact object (presumably an accretion disk) can
actually dominate the total output. Recent {\em RXTE} observations
have shown that \object{LMC X-3}, like most black-hole candidates (but
not \object{LMC X-1}), oscillates between X-ray states, characterized
by a softer spectrum when the luminosity is higher \cite{wil01}. Our
spectroscopy was taken when the X-ray luminosity was relatively high,
according to the {\em RXTE}/ASM quicklook results.

\begin{figure*}
\begin{picture}(500,240)
\put(0,0){\includegraphics{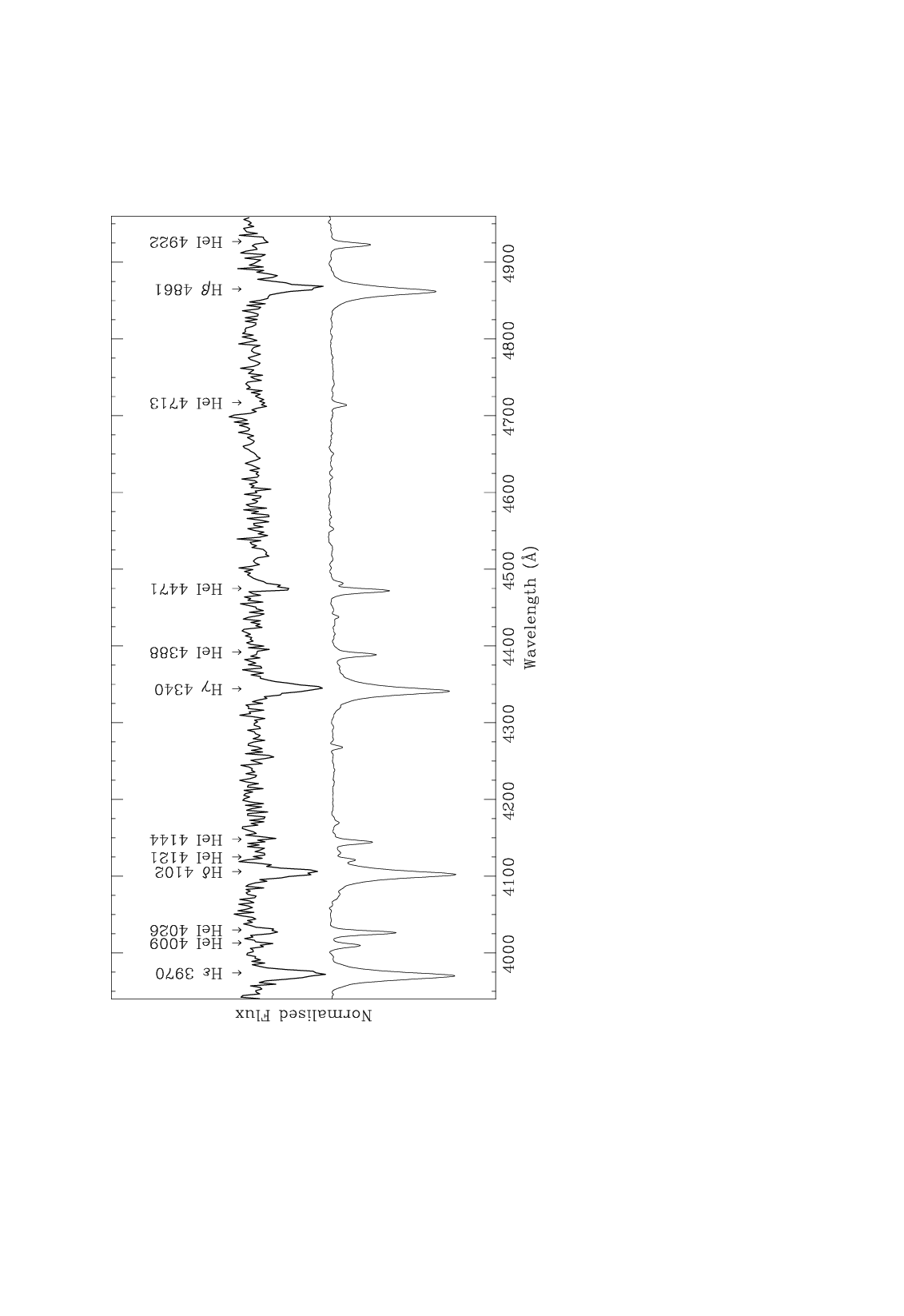}}
\end{picture}
\caption{Blue spectrum of the optical counterpart to \object{LMC X-3}
  (top). The bottom spectrum is that of the Galactic MK B2V standard  
  \object{22 Sco}. Note the weakness of the Balmer lines in the spectrum of
  \object{LMC X-3}, suggesting that there is a significant
  contribution to the continuum from another source (the accretion disk). }
\label{fig:weako}
\end{figure*}

The spectrum of \object{LMC X-3} is displayed in Fig.~\ref{fig:weako},
together with that of the B2V MK standard 
\object{22 Sco}. As remarked by previous
authors, the intensity of the Balmer lines in the spectrum of
\object{LMC X-3} is much smaller than expected, presumably due to the
fact that the accretion disk around the black hole contributes a
substantial fraction of the continuum (as is expected from the
photometric variations). Some \ion{He}{i} lines are comparatively very
strong, specially \ion{He}{i}~$\lambda$4471~\AA\ and other triplet
lines. Their strength is absolutely inconsistent with a B5
classification. Their ratios with respect to the Balmer line suggest a
spectral type even slightly earlier than the B3V proposed by Cowley et
al. (1983). However, it is clear that the spectrum is not that of a
normal star. We observe broad emission corresponding to
\ion{He}{ii}~$\lambda$4686~\AA\ and perhaps the Bowen blend. This
emission, which must originate from the accretion disk, has never been
reported before. Even more puzzling are the strong absorption features
at $\sim \lambda$4260\AA\ and $\sim \lambda$4520\AA, which do not
correspond to any lines generally seen at this spectral type.

Therefore it is difficult to assign a spectral type to the optical
counterpart to \object{LMC X-3}. Based on the strength of the
\ion{He}{i} lines, the spectral type should be around B2.5V, but we
doubt that such a spectral classification is really reflecting the
properties of the star. We suspect that the strength of these lines
must be reflecting an overabundance of He in the outer layers of the
star.

\subsection{\object{CAL E}}

The X-ray source \object{RX J0502.9$-$6626} was originally detected by
the {\em Einstein} observatory \cite{cow84} at a flux of $\sim
3\times10^{36}\:{\rm erg}\,{\rm s}^{-1}$. The source was detected
three times with the {\em ROSAT} PSPC at luminosities $\sim
10^{35}-10^{36} \:{\rm erg}\,{\rm s}^{-1}$ and once with the HRI
during a bright outburst at $\approx 4\times10^{37}\:{\rm erg}\,{\rm
  s}^{-1}$ \cite{sch95}. During the outburst, pulsations at $P_{{\rm
    s}}=4.0635\:{\rm s}$ were detected. The identification of this
source with the Be star \object{[W63b]564} = 
\object{EQ 050246.6$-$663032.4}  \cite{woo63} was
confirmed by Schmidtke et al. (1994).

The spectrum of the optical counterpart to \object{CAL E} is displayed
in Fig.~\ref{fig:twocals}. The obvious presence of
\ion{He}{ii}~$\lambda$4686\AA\ and absence of
\ion{He}{ii}~$\lambda$4200\AA\ indicate that the spectral type is close
to B0. If the object is on the main sequence, the strength
of \ion{Si}{iv}~$\lambda$4089\AA\ indicates a spectral type B0.2 or
earlier. Since \ion{C}{iii}~$\lambda$4650\AA\ $\ga$
\ion{He}{ii}~$\lambda$4686\AA\ and \ion{He}{ii}~$\lambda$4200\AA\ is
not seen, the object cannot be earlier than B0. If it is a giant, it
may be slightly earlier, which would be supported by the lack of  
\ion{Si}{iii} features. Therefore the spectral type is constrained to
be B0V or O9.5-B0III

For these spectral types, $(B-V)_{0}\approx-0.26$ (Wegner 1994).
 The source was repeatedly observed 
by Schmidtke et al. (1996), who found very
significant variability by over $0.4\:{\rm mag}$. The changes in $V$
were well correlated with changes in the $(B-V)$ colour, as is typical
in Be stars. The fainter points measured (in February 1992) were
$V\approx14.28$ and $(B-V)=-0.10$, which are very close to our
 measurements in October 1996. Assuming that the interstellar
reddening is $E(B-V)=0.10$, the derived intrinsic magnitude is
$M_{V}=-4.2$, typical of a B0V star. If all the reddening is
 interstellar, $M_{V}=-4.5$, also compatible with the spectral type.  

\subsection{\object{CAL 9}}

This {\em Einstein} and {\em ROSAT} variable source was identified
with a Be star by Schmidtke et al. (1994). Later Schmidtke et
al. (1996) identified this star with
\object{HV 2289}, a known variable with a large amplitude of
variability. Photometry by Schmidtke et al. (1996) showed little
variability over more than one year in $1991-1993$. The faintest
points corresponded to $V=14.5$, $(B-V)=-0.10$.

\begin{figure*}
\begin{picture}(500,240)
\put(0,0){\includegraphics{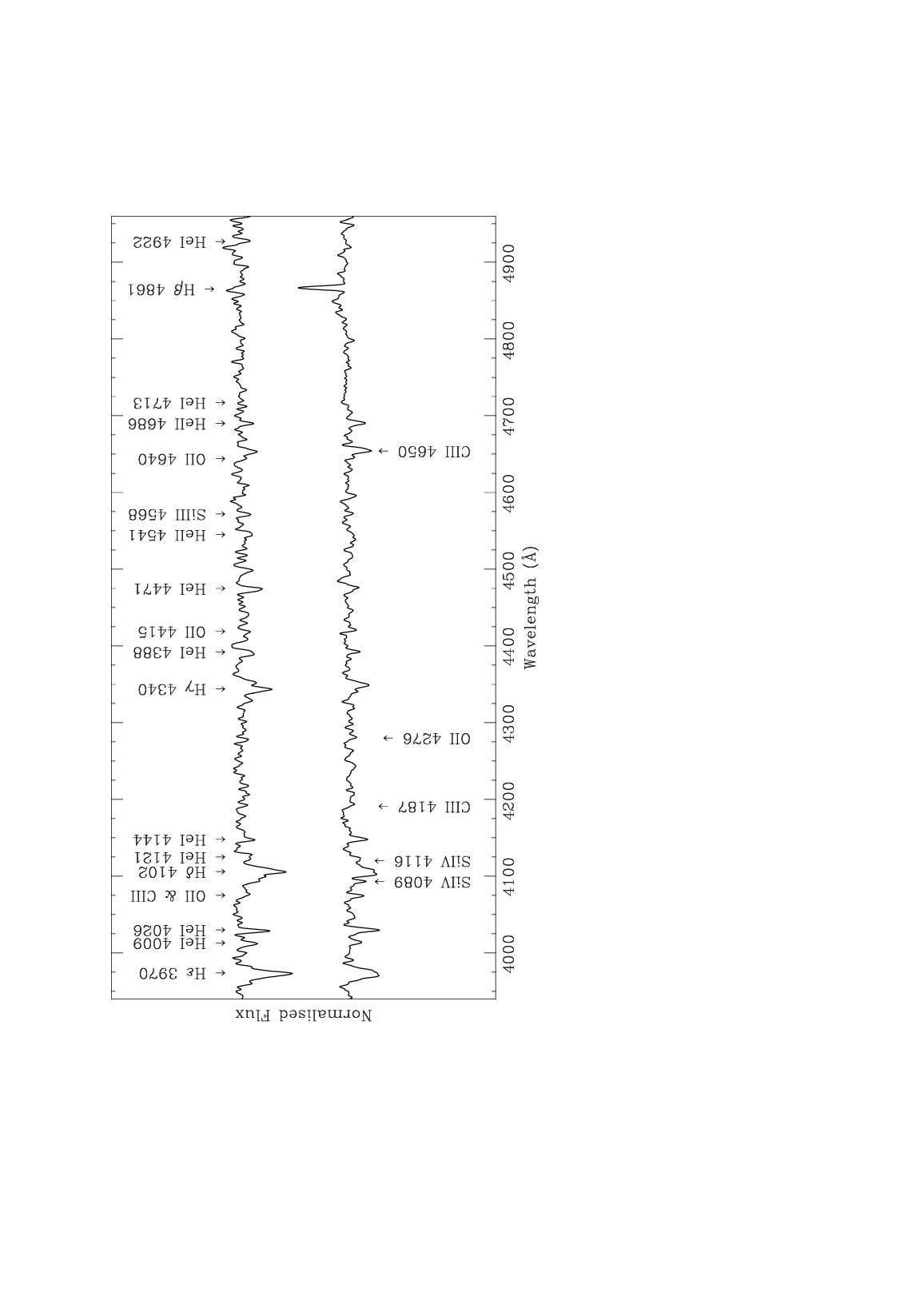}}
\end{picture}
\caption{Blue spectra of the optical counterparts to \object{CAL E} 
  (bottom) and \object{CAL 9} (top). The spectra have been smoothed
  with a $\sigma=1.4$ Gaussian for display. The absorption feature at
  $\sim \lambda4510$~\AA\ in the spectrum of \object{CAL 9} appears
  only in one of the summed spectra and is likely an artifact.}
\label{fig:twocals}
\end{figure*}

We have two spectra of this source, one from November 1999 and a
second one from October 2000. The emission components of the Balmer
and \ion{He}{i} lines increased considerably during the interval
between the two spectra. H$\delta$, which was basically in absorption
in 1999 shows substantial in-filling in 2000. Our photometric data,
taken in late 1996, show the star slightly brighter (though rather 
redder) than at minimum. All this suggests that the star was in a low 
state from 1995 to 1999, with the circumstellar disk almost absent and
has only recently come back to a Be phase.

The spectrum  displayed in Fig.~\ref{fig:twocals} is a sum of our two
spectra binned to the same resolution.
Again we observe weak \ion{He}{ii} lines indicating a spectral type
close to B0. The absence of \ion{He}{ii}~$\lambda$4200\AA\ indicates
that it is not an O-type star. Again the condition
\ion{He}{ii}~$\lambda$4686\AA\ $\simeq$ \ion{C}{iv}~$\lambda$4650\AA\
supports a spectral type B0-B0.2. Assuming that the faintest points in
the photometry of Schmidtke et al. (1996) represent values close to
those intrinsic to the star, we find $E(B-V)=0.16$ and hence
$M_{V}=-4.2$. If the interstellar reddening is lower
$E(B-V)=0.08$, $M_{V}=-4.0$. This values are typical of a B0V star.

\subsection{\object{RX J0520.5$-$6932}}

This X-ray source has only been seen once at a low X-ray
luminosity (Schmidtke et al. 1994). The lightcurve of the optical
counterpart, however, 
exhibits significant modulation with a period of $24.5\:\mathrm{d}$,
which is interpreted as the orbital period (\cite{coe01}). The
high-resolution spectrum of this source from November 1999 was
presented in Coe et al. (2001), where an approximate O9V spectral type
was derived. Since the SNR of the spectrum was relatively low, we took
two further spectra using the lower-resolution grating \#33 in
September 2000 and October 2000. When the three spectra are compared,
obvious variability in the emission components is seen, with the
photospheric component of the Balmer lines becoming progressively
deeper as emission disappears. This suggests that the circumstellar
envelope is in the process of dispersing.

\begin{figure*}
\begin{picture}(500,240)
\put(0,0){\includegraphics{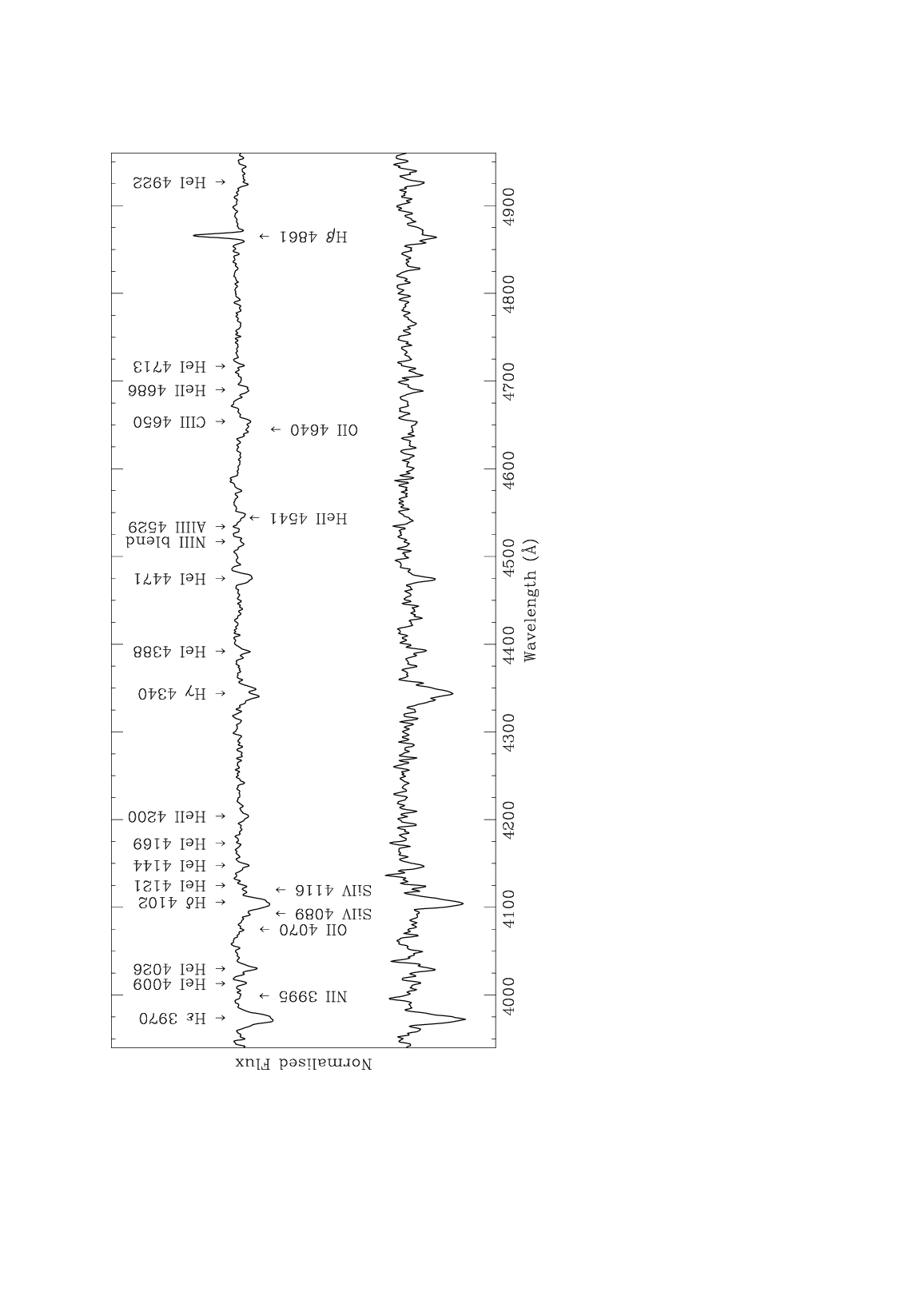}}
\end{picture}
\caption{Summed spectra of the optical counterparts to \object{RX
    J0520.5$-$6932} (top) and \object{RX J0544.1$-$7100} (bottom).}
\label{fig:ogle}
\end{figure*}

Figure~\ref{fig:ogle} shows the sum of the three spectra, binned to the
same resolution. The strengths of
\ion{He}{ii}~$\lambda\lambda$4200,4541\AA\ and the weakness of the
\ion{Si}{iii} lines confirm a spectral type close to O9. This is also
in agreement with the relative strengths of the \ion{Si}{iv} lines
(now visible on the wings of H$\delta$) and nearby \ion{He}{i} lines
-- if the latter are not affected by emission. At this spectral type, 
\ion{He}{ii}~$\lambda$4686\AA\ is expected to be stronger than 
\ion{C}{iii}~$\lambda$4650\AA\ for the main sequence. The weakness of
the \ion{He}{ii} line suggests a higher luminosity. From the
photometric data in Coe et al. (2001), with the derived
$E(B-V)\approx0.3$, the intrinsic magnitude is $M_{V}=-4.7$. However,
the average OGLE photometric values in Coe et al. (2001) imply
$E(B-V)=0.22$. Since there
is certainly some circumstellar contribution to this average $V$
(because the source is periodically variable), the derived
$M_{V}=-4.5$ has to be taken as an upper limit. Therefore the
photometric data support a main-sequence classification. We adopt 
O9V.

\subsection{\object{RX J0544.1$-$7100}}

This source, identified with \object{1SAX J0544.1$-$710}, is a
transient X-ray pulsar ($P_\mathrm{s}=96\,\mathrm{s}$) with the hardest
X-ray spectrum observed by {\em ROSAT} in the LMC (Haberl \& Pietsch
1999). As in the case of
\object{RX J0520.5$-$6932}, observations of the
optical counterpart were presented by Coe et al. (2001), who found it
to display large variability in the $I$-band lightcurve and H$\alpha$
in emission. An approximate spectral type of B0V was proposed, based
on the spectrum obtained for this campaign in November 1999. Again, we
have taken a second lower-resolution spectrum of this source in
October 2000. 

Figure~\ref{fig:ogle} shows the sum of the two spectra, binned to the
same resolution. The strengths of \ion{He}{ii}~$\lambda\lambda$4541,
4686\AA\ are compatible with a B0V spectral type, but 
\ion{C}{iii}~$\lambda$4650\AA\ is far too weak, even for an LMC
source. This could be related to the presence of the strong line at $\sim
4045$\AA\ (real, since it is present in both spectra), which would then be
a \ion{N}{ii} blend, suggesting some N enhancement. 
Some peculiarity, however, remains, since the
features at $\lambda 4210$\AA\ and $\lambda 4829$\AA\, which we cannot
identify, also seem to be real. Higher resolution observations of
this object are necessary

This star is photometrically very variable. The OGLE $I$-band
lightcurve \cite{coe01} shows variations of $\sim 0.3\:{\rm
  mag}$. The photometry from January 1999 indicates $(B-V)=0.1$
implying $E(B-V)=0.36$. Such value is rather high to be purely
interstellar and again indicates some contribution from the
circumstellar disk. Taking the interstellar contribution to be within
$E_{{\rm is}} = 0.10$ and $E_{{\rm is}} = 0.30$, we obtain $M_{V}$ in
the range $-3.3$ to $-3.9$, which correspond to a dwarf in the
B0\,--\,B1 range.

\subsection{\object{RX J0529.8$-$6556}}

The transient X-ray source \object{RX J0529.8-6556} was detected
during one single outburst as a 69.5-s X-ray pulsar by Haberl et
al. (1997), who identified it with a relatively bright blue star
showing weak H$\alpha$ emission.

\begin{figure*}
\begin{picture}(500,240)
\put(0,0){\includegraphics{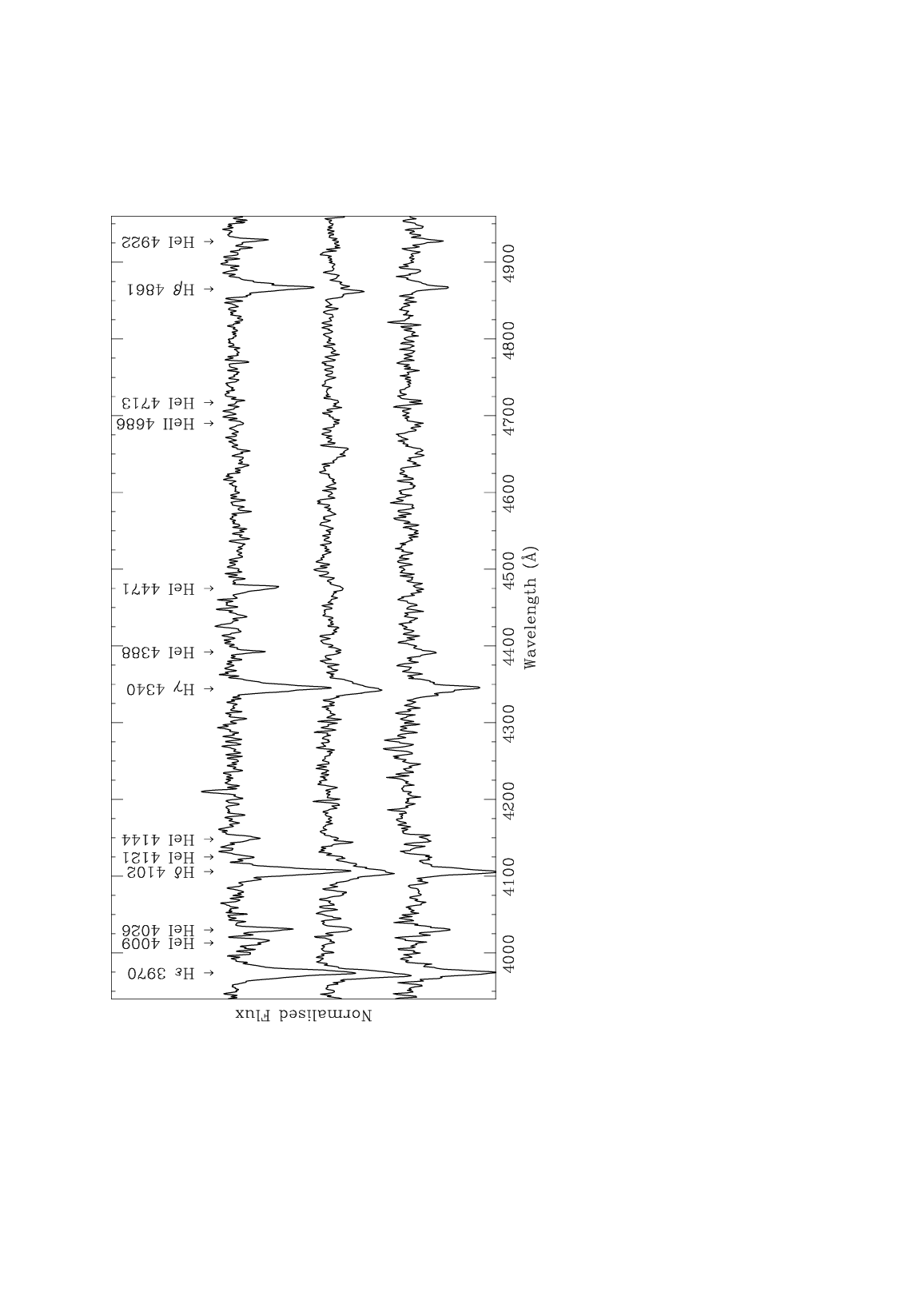}}
\end{picture}
\caption{Blue spectra of the optical counterparts to \object{RX
    J0529.8$-$6556} (top), \object{EXO\,0531.1$-$6609} (middle) and
 \object{H0544$-$665} (bottom).} 
\label{fig:others}
\end{figure*}

The spectrum of the optical counterpart is displayed in
Fig.~\ref{fig:others}. There is little emission in the Balmer
lines, but \ion{He}{i}~$\lambda$4713\AA\ is hardly
seen. \ion{He}{ii}~$\lambda$4686\AA\ is very weak suggesting a
spectral type later than B0. The ratios of the \ion{He}{i} lines are
compatible with a spectral type close to B0.5. The presence of
\ion{Si}{iv}~$\lambda$4089\AA\ is incompatible with a later type. All
the metallic lines are very weak and the Bowen blend is hardly
seen. The spectral type is B0.2 or B0.5, with the weakness of the
metallic lines making a main sequence classification much more
likely. 

Our photometry indicates $(B-V)=-0.16$. The $E(B-V)=0.08$ is
consistent solely with interstellar absorption, which is expected,
given the absence of emission in the Balmer lines. The derived
$M_{V}=-3.7$ is consistent with a B0.5V star.

\subsection{\object{EXO\,0531.1$-$6609}}

This X-ray source was discovered by {\em EXOSAT} during an outburst in
October\,-\,November 1983 \cite{pak85}. Later it was detected by the
X-ray telescope on Spacelab 2 in July/August 1985
at a flux $\sim 10^{37}\:{\rm erg}\,{\rm s}^{-1}$
\cite{han89}. Monitoring with {\em ROSAT} confirmed it to be highly
variable. The source was detected at a low level in November 1991 and
then during a bright outburst in April 1993 \cite{hab95b}. During the
outburst, pulsations were detected at $P_{\rm s}=13.7\:{\rm s}$. From
a study of the variability of the pulsations, Dennerl
 et al. (1996) derive a likely (though not certain) orbital solution
 with $P_{{\rm orb}}=24.5\:{\rm d}$ and low eccentricity. All the
 outbursts detected from \object{EXO\,0531.1$-$6609} have been rather
longer than the proposed orbital period, suggesting that they are giant 
outbursts (see Negueruela 1998).

The counterpart is a relatively faint Be star, the Northern component
of a very close double. The second star is slightly fainter, but we 
have not managed to disentangle their contributions in any of our
images and cannot therefore derive reliable photometry. Given the
rather poor seeing during our spectroscopic observations, even though
our spectra were taken with the slit set so as to leave out the
companion, we expect some contamination. In any case, spectra taken in
1999 and 2000 do not differ significantly, except for some variability
in the in-filling of the cores of Balmer lines.
The summed spectrum  is displayed in
Fig.~\ref{fig:others}. \ion{He}{ii}~$\lambda$4686\AA\ may be very weakly
present, indicating a spectral type later than B0. The relatively
strong \ion{C}{iii}~$\lambda$4650~\AA\ line makes it B1 or earlier. 
Though several weak lines
attributable to \ion{O}{ii} are visible, the \ion{Si}{iii} triplet is
not visible, suggesting main sequence. This is supported by the broad
Balmer lines (except H$\beta$, which is filled in). A spectral type
around B0.7V seems likely.

\subsection{\object{H0544$-$665}}

\object{H0544$-$665} was discovered as a weak X-ray source by Johnston
et al. (1979) using the {\em HEAO-1} scanning modulation
collimator. Its luminosity $\sim 
10^{37}\:\mathrm{erg}\,\mathrm{s}^{-1}$ is typical of a Be/X-ray
binary in outburst. In the rather large ($r = 1\arcmin$) error box,
there is only one object bright enough to be an early-type Be star in
the LMC. Based on its large optical variability, van der Klis et
al. (1983) proposed it as the  optical counterpart, though they did
not detect any emission in their spectra. Later Stevens et
al. (1999) detected H$\alpha$ and
H$\beta$ in emission in February 1998.

Our spectrum of this source, taken in November 1999 is shown in 
Fig.~\ref{fig:others}. The spectrum is rather noisy, due to the
relative faintness of the source, but weak \ion{He}{ii}
$\lambda\lambda$ 4541, 4686~\AA\ lines are
visible, suggesting a spectral type B0V. 

van der Klis et al. (1983)
obtained extensive photometry of the source and found that there was a
correlation between its brightness and the $(B-V)$ colour, as is
typical of Be stars. From their data, some variability in $B$ may be
deduced, up to 0.2 mag. Their faintest (and therefore bluest)
datapoints imply $V\approx 15.5$, $(B-V)\approx -0.20$. This is
consistent with the intrinsic colour of a B0V star reddened with the
typical $E(B-V)\approx0.07$ for LMC sources. However these magnitudes
imply $M_{V}=-3.0$, which is too faint for a B0V star and more in line
with B1-B.15V. Since the spectrum seems to be that of a B0V star, we
must conclude that the star is underluminous for its spectral type 
($M_{V}=-4.2$ after Vacca et al. 1996).

\subsection{\object{1A\,0535$-$66}}

Probably the best known and least understood Be/X-ray transient,
\object{1A\,0535$-$66} was discovered by the {\em Ariel 5} satellite
in June 1977, during at outburst in which the flux peaked at 
$\sim 9\times10^{38}\:\mathrm{erg}\,\mathrm{s}^{-1}$ \cite{wc78}. When
active, \object{1A\,0535$-$66} displays very bright short X-ray outbursts
separated by 16.6 days, which is believed to be the orbital
period. The optical counterpart experiences drastic changes in the
spectrum, with the appearance of strong P-Cygni-like emission lines,
and brightening by more than 2 mag in the $V$ band
\cite{cha83}. Detection of a 69-ms pulsation in the X-ray signal has
been reported only once \cite{ski82}.

\begin{figure*}
\begin{picture}(500,240)
\put(0,0){\includegraphics{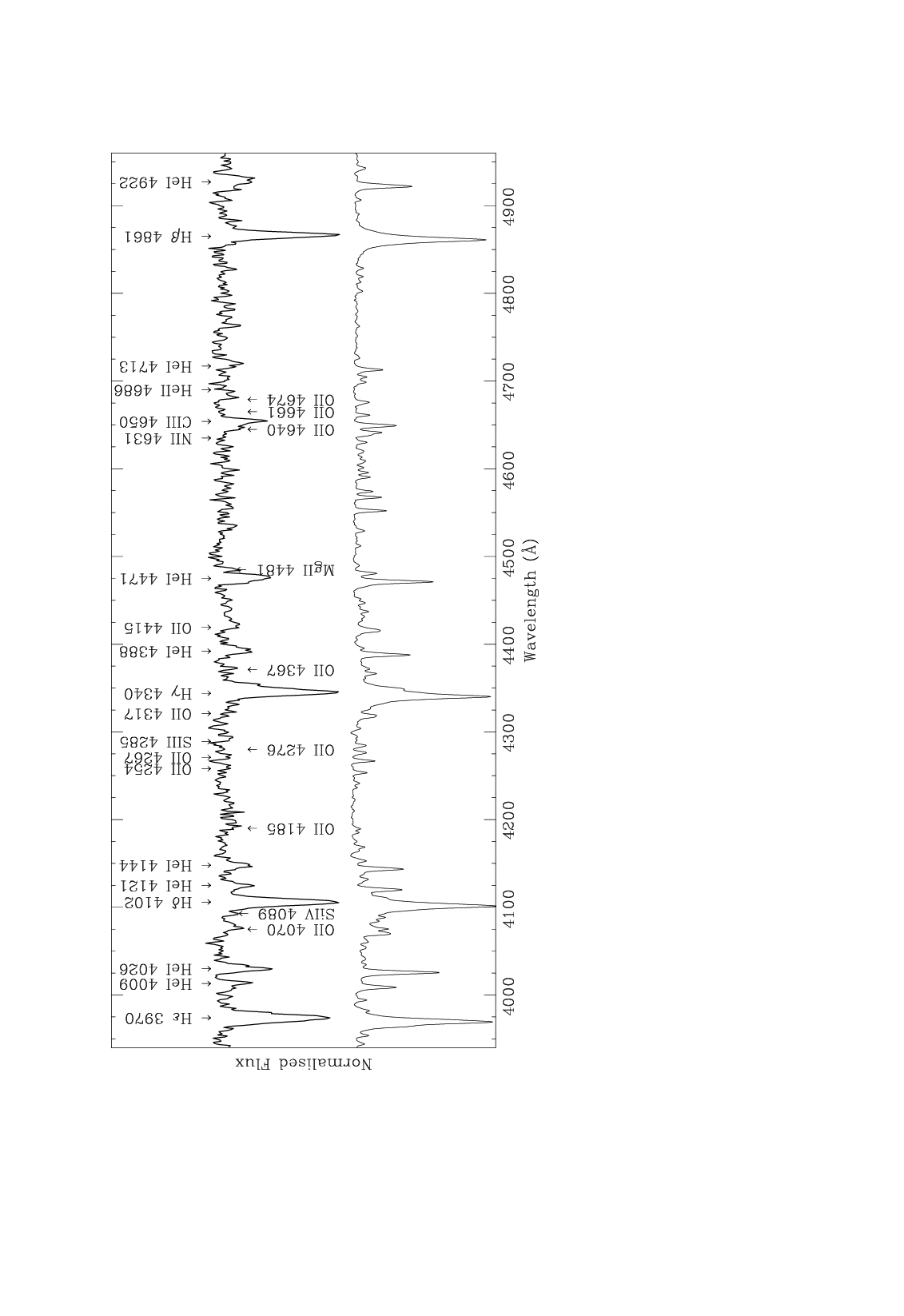}}
\end{picture}
\caption{The spectrum of the optical counterpart to
  \object{1A\,0535$-$66} (top) compared to the B1III standard 
\object{HD 23180}.}
\label{fig:mega}
\end{figure*}

Bright X-ray outbursts from \object{1A\,0535$-$66} have not been seen
since 1983 but some lower level activity has always been
observed. Monitoring in the optical between 1993 and 1998 has shown
recurrent outbursts, when the source brightens by up to $\sim 0.6\,{\rm
  mag}$ in $V$, modulated at a period of $16.651\,{\rm d}$, which is
identified with the period seen in the X-rays \cite{alc01}. In
addition, there is a longer-term modulation with period
$P=420.8\pm0.8\,{\rm d}$, corresponding to smoother changes with an
amplitude of $\approx 0.4\,{\rm mag}$. Apparently, the short outbursts
only occur during the phase of the 421-d period in which the source is
faint \cite{alc01}.

There is very little difference in the photospheric features  between
the October 1999 and September 2000 spectra, which correspond to
approximate phases $\phi=0.89$ and $\phi=0.65$ in the 421-d period as
defined by Alcock et al. (2001). Some weak emission can
be present in the bottom of some \ion{He}{i} lines in the 1999
spectrum. For this reason, we display the 2000 spectrum of the optical
counterpart in Fig.~\ref{fig:mega} together with the B1III MK standard
\object{HD 23180}. The overall appearance of both spectra is very
similar. The \ion{Si}{iii} lines are weaker in the spectrum of 
\object{1A\,0535$-$66}, as corresponds to an LMC source. However, the
\ion{O}{ii} blend at $\lambda4650$\AA\ and
\ion{Mg}{ii}~$\lambda$4481\AA\ appear stronger than in the Galactic
standard, which together with the smaller intensity of H and
\ion{He}{i} lines would suggest a higher luminosity. In this case, the
spectral type of \object{1A\,0535$-$66} in quiescence would be B1II. 

Such a high luminosity would result in a very extended atmosphere,
suggesting that \object{1A\,0535$-$66} may indeed be close to
overflowing  its Roche lobe at periastron. 
However, we point out that a weak feature that could be
\ion{He}{ii}~$\lambda$4686\AA\ is visible in our 2000 spectrum. This
feature is not seen in 1999, but 
\ion{He}{ii}~$\lambda$4686\AA\ emission has been observed in
\object{1A\,0535$-$66} even during off 
states \cite{hut85}. If some emission was occulting a very
weak \ion{He}{ii}~$\lambda$4686\AA\ photospheric feature, then the
strength of \ion{Si}{iv}~$\lambda$4089\AA\ would support a B0.5
classification. The line at $\lambda4650$\AA\ would then be a blend of
\ion{O}{ii} and \ion{C}{iii}, making the classification B0.5III.

Our spectral classification for \object{1A\,0535$-$66} as B0.5III is
rather different from that of Charles et al. (1983), B2IV, but in good
agreement with the estimate from a quiescent spectrum of Pakull \&
Parmar (1981). The colours of  \object{1A\,0535$-$66} have been
observed to change significantly. Pakull \& Parmar (1981) report a
$(B-V)=-0.20$ during February-April 1980, when the source was in the
``off'' state. Such value must be the intrinsic colour of the star,
since it implies $E(B-V)=0.04$. The corresponding $V=14.83$ indicates
then $M_{V}=-3.5$, which may be compatible with a main-sequence star,
but certainly not with a giant. Alcock et al. (2001) have suggested
that the actual quiescent magnitude of \object{1A\,0535$-$66} is
$V\approx14.4$ and that the dips at $V\approx14.8$ are caused by the
development of an obscuring shell. Even then $M_{V}=-4.0$ would be
rather lower than expected for a giant star.

\subsection{\object{RX J0531.5$-$6518}}

This source was detected with the {\em ROSAT} PSPC in June 1990
\cite{hap99}. The source is probably variable, since other pointings
failed to detect it. The error circle shown by Haberl \& Pietsch
(1999) contains only one relatively bright object. The
spectrum of this star taken in November 1999 seems typical of a normal
early-type B star. The spectrum from October 2000 shows substantial
infilling of H$\beta$ and a decrease in the intensity of all Balmer
lines. Therefore this object is probably a Be star coming back from an
extended disk-less phase and so the correct optical counterpart.

The summed spectrum of the star is displayed in
Fig.~\ref{fig:maybes}. No \ion{He}{ii} lines are visible and
the Bowen blend seems to be absent. The absence of any \ion{Si}{iii}
lines, the ratios of the \ion{He}{i}
lines and the strength of \ion{Mg}{ii}~$\lambda$4481\AA\ indicate a
spectral type B2. However, the presence of some \ion{O}{ii} lines is
surprising, especially because the line tentatively identified as
\ion{N}{ii}~$\lambda$3995~\AA\ would be indicating N enhancement. 
We provisionally adopt a spectral type B2V, though some anomaly seems
to be present.

Our photometry gives $(B-V)=-0.16\pm0.03$, implying a reddening
$E(B-V)=0.05$, consistent with interstellar. This again favours the
idea that the object was not in a Be phase during 1999. The derived
$M_{V}=-2.4$ is consistent with a B2V spectral type (Schmidt-Kaler
1982).

\subsection{\object{RX J0535.0$-$6700}}

The variable source \object{RX J0535.0$-$6700} was observed by the
{\em ROSAT} PSPC at a luminosity $\sim
3\times10^{35}\:\mathrm{erg}\,\mathrm{s}^{-1}$ \cite{hap99}. Its
positional coincidence with an optically variable star in the LMC
(\object{RGC28} in Reid et al. 1988) is very good. The star displays
periodic variability in its $I$-band lightcurve at $P=241\,{\rm d}$,
which Reid et al. (1998) originally believed to be the period of a
Mira variable. Since the optical counterpart to \object{\,0535$-$66}
also appears in this catalogue, mistakenly taken by a Mira variable,
Haberl \& Pietsch (1999) suggest that \object{RGC28}
might be a Be star, in which case it would be the optical
counterpart and the variability would be related to the orbital
period. 

\begin{figure*}
\begin{picture}(500,240)
\put(0,0){\includegraphics{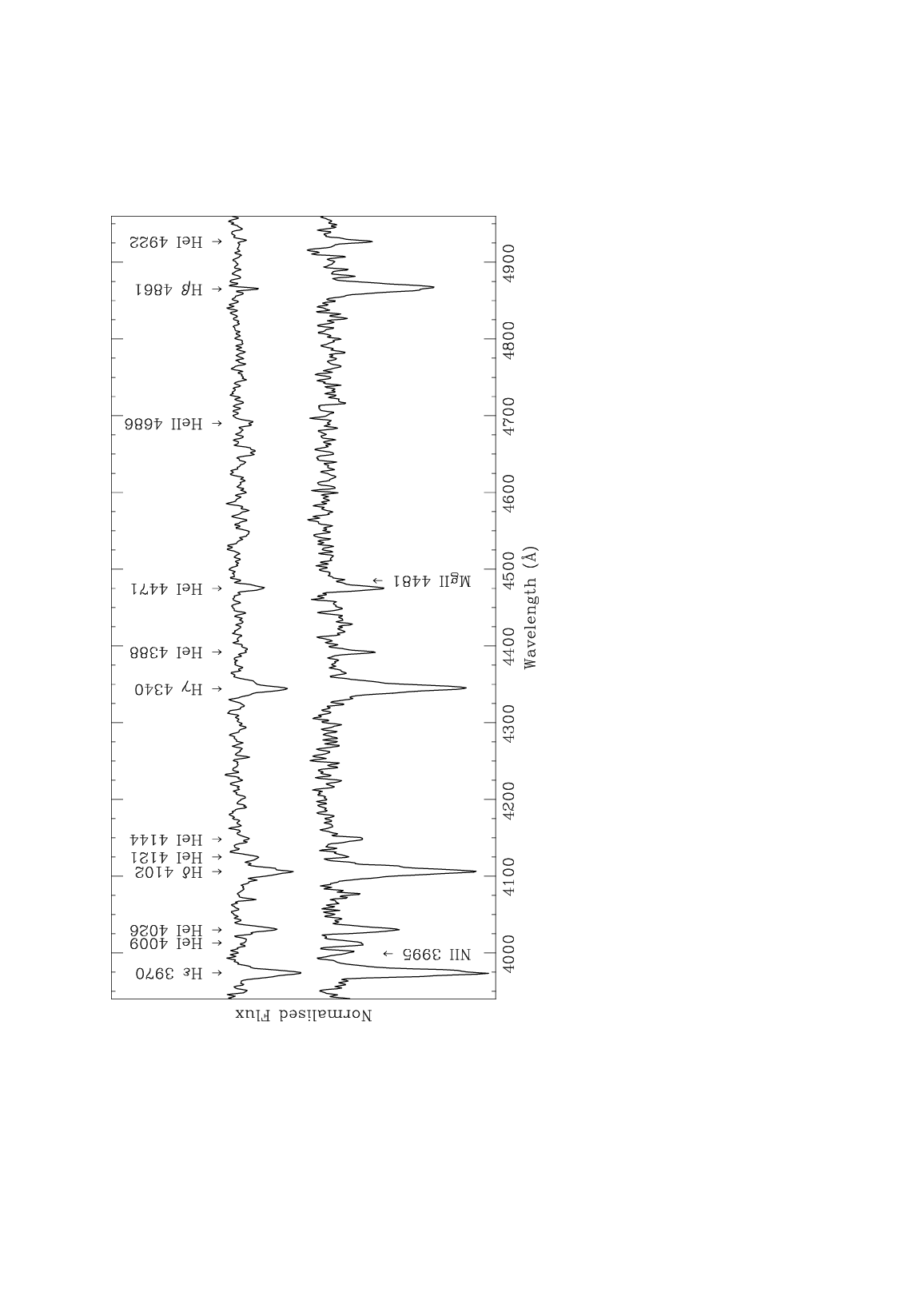}}
\end{picture}
\caption{Blue spectra of the proposed optical counterparts to 
    \object{RX J0531.5$-$6518} (bottom) and \object{RX J0535.0$-$6700} 
(top).} 
\label{fig:maybes}
\end{figure*}

The spectrum of \object{RGC28} is displayed in
Fig~\ref{fig:maybes}. The object is obviously an early-type Be star,
confirming the suspicion of Haberl \& Pietsch (1999), and likely the
optical counterpart to \object{RX J0535.0$-$6700}. H$\beta$ is almost
completely filled in by emission. The presence of
\ion{He}{ii}~$\lambda\lambda$4200, 4541\AA\ indicates a spectral type
B0 or earlier. If the object was a Galactic source, the ratio 
\ion{C}{iii}~$\lambda$4650\AA $\simeq$ \ion{He}{ii}~$\lambda$4686\AA\
would indicate a B0V star. In the LMC, however, the relatively strong
metallic spectrum could indicate a slightly higher luminosity and
therefore slightly earlier spectral type, making O9.5III in principle also
compatible with the observations.

Our photometry indicates $(B-V)=-0.07$. The interstellar reddening
is then at most $E(B-V)=-0.19$. This implies $M_{V}=-4.0$, indicating
the main sequence classification. Therefore we adopt B0V.

\section{Discussion}
\label{sec:dis}

We have obtained accurate spectral classifications for an almost
complete sample of HMXBs in the LMC.
Among these objects, there are three well-known bright persistent
X-ray sources, two of which are considered to 
be black hole candidates (LMC X-1, O8III-V; LMC X-3, B2?Vp), while the
third is a bright X-ray pulsar fed by Roche-Lobe overflow from the
O8III companion (LMC X-4). In addition, we have found that the
optical counterpart to \object{RX J0532.5$-$6551} has a spectral type
B0II and therefore this source is likely to be a wind-fed accreting
system, like Vela X-1. 
All the other sources in the sample are likely Be/X-ray
binaries. Eight of them had been considered as such in previous work
(of which, 5 are X-ray
pulsars) and two further suggestions seem to be confirmed by our
results. The list of confirmed and likely Be/X-ray binaries in the LMC
is summarized in Table~\ref{tab:bex}, while the list of possible HMXBs
which have not been included in this investigation is given in
Table~\ref{tab:unknowns}. 

Since we have relied on the derived absolute magnitudes of our objects
to determine their luminosity classes, the choice of a particular
calibration has a direct bearing on the results. Throughout
Section~\ref{sec:spec}, we have used the calibration of Vacca et
al. (1996) for O-type stars, complemented by that of Schmidt-Kaler
(1982) for early B stars. By using this calibrations, we have found
that three objects seem to be underluminous for their spectral type:
\object{LMC X-4} is moderately underluminous for O8V, though a
spectral type is difficult to assign. On the other hand, both 
\object{1A\,0538$-$66} and \object{H0544$-$665} have well defined
spectral types and their absolute magnitudes seem to be slightly lower
than expected.

The distance modulus to the LMC used through this work is
$(M-m)_{0}=18.24$, after Udalsky (2000). The validity of this value is
still open to discussion, with many authors supporting a value
$(M-m)_{0}=18.5$. We just note here that the distance modulus adopted
does not affect in any significant way the results obtained. The only
changes that would result from adopting the ``longer'' distance would
be that the intrinsic luminosity of the optical companion of 
\object{LMC X-4} would become
marginally compatible with an O8V spectral type (but still not with
O8III) and that the intrinsic luminosity of 
\object{EQ 050246.6$-$663032.4} (CAL E) would
then be compatible with a B0III spectral type. \object{H0544$-$665} and
\object{1A\,0538$-$66} would still be underluminous for their spectral
types. 

A new  calibration based on Hipparcos distances has been derived by 
Wegner (2000), who finds on average rather lower absolute magnitudes
at a given spectral type than previously assumed (typically by $\sim
1$ mag). By using this
calibration, we could solve the problem of our three underluminous
objects, but we would create a larger one, since the magnitudes that
we derive (typically $M_{V}\approx -4$ for B0V objects) would then
mean that all the other counterparts are bright giants. 
Therefore we conclude that our data support the older
``brighter'' calibrations, even though we have used a ``short''
distance for the LMC.

\subsection{A preliminary comparison to the Galactic population}

Our sample of the LMC HMXB
population is complete, in the sense that we have observed all the
systems with confirmed optical counterparts. As shown in
Table~\ref{tab:unknowns}, only two possible Be/X-ray binaries are left
out (\object{RX J0516.0$-$6916} and \object{RX J0532.4$-$6535}). 
The main limitation to any conclusions to be drawn from this
sample (i.e., how representative it is?) stems from our inability
to judge how many other X-ray sources remain undiscovered. In this
respect, it is clear that, unlike in the Galaxy, we are not missing
any sources because they are very absorbed. The {\em ROSAT} PSPC
pointed at most locations in the LMC more than once, with many fields
being observed close to ten times \cite{hap99}. Since the sensitivity
limit of {\em ROSAT} allowed the detection of relatively weak LMC sources, 
only very weak persistent X-ray
sources (with $L_{\rm x}\la10^{34}\:{\rm erg}\,{\rm s}^{-1}$) may have
been missed, specially in fields where no deep exposures have been
carried out.

The main uncertainty is therefore what fraction of transient sources
has not been discovered yet because they were not active when they
were looked at. There is no obvious way to assess this number, but a
comparison with the Galactic sources may indicate that the census is
still relatively incomplete, since a large fraction of Be/X-ray
transients are only active as X-ray sources for relatively short
periods. 

At first sight, the LMC HXMB population is not significantly different
from the Galactic one. All the LMC systems seem to have Galactic
equivalents. \object{LMC X-4} is similar to the Roche-Lobe
overflow accreting pulsar \object{Cen X-3}. \object{LMC X-1} would be
its equivalent with a black hole companion, not too different from
\object{Cyg X-1}, except in the mass transfer mechanism.
\object{LMC X-3} is similar to the recently found black hole candidate
\object{SAX J1819.3$-$2525}/\object{V4641 Sgr} for whose components
(B9III+BH) Orosz et al. (2001)
derive masses which are quite close to those estimated for \object{LMC
X-3}. The fact that \object{SAX J1819.3$-$2525} is a transient X-ray source,
while \object{LMC X-3} is persistent, may be related to the smaller orbit of
\object{LMC X-3}. 

Among the list of known and probable HMXBs of Liu et al. (2000) we
find that there are 40 Galactic systems with identified optical
counterparts -- leaving aside some objects like \object{$\gamma$ Cas},
\object{1E\,1024.0$-$5732} or \object{1H\,0521+373} whose nature as
HMXBs is unclear. Of these, 23 are BeXBs and 10 are SXBs. 7 objects
are not included in any of the two groups. In particular, 
we exclude the binaries \object{1E\,0236.6+6100} and \object{SAX
  J0635+0533} from the count of Be/X-ray binaries, because they are
not believed to be accretion driven. We have also excluded \object{XTE
  J0421+560}, because the exact nature of its two components is still
unknown. The respective fractions are then 25\%
SXBs and 58\% BeXBs. Adding objects without optical
counterparts which are believed to belong to one of the two groups
because of their X-ray properties or orbital solutions gives 13 SXBs
and 37 BeXBs. The fraction SXBs/BeXBs among systems with identified
counterparts is 0.43, which reduces to 0.35 when candidate
unidentified sources are added.
 
Of the 10 Galactic SXBs (and 13 likely SXBs), only one (\object{Cen X-3}) is
powered by Roche-lobe overflow (RLO). All the others are wind-fed
systems. Moreover, three of the Galactic HMXBs that have not been included in
any of the two groups are also wind-fed X-ray sources: \object{RX
  J1826.2$-$1450} and \object{4U\,2206+54} contain main-sequence
O-type stars (c.f. Negueruela \& Reig 2001) and \object{Cyg X-3}
likely contains a Wolf-Rayet star. This means that we know of 16
wind-fed X-ray sources in the Galaxy (most of them persistent and with
$L_{\rm x}\ga10^{35}\:{\rm erg}\,{\rm s}^{-1}$) against 1 (or maybe 2
if \object{RX J0541.4$-$6936} really belongs to this category) such
sources in the LMC. Since some wind-fed systems are weaker X-ray
sources ($L_{\rm x}\approx10^{34}\:{\rm erg}\,{\rm s}^{-1}$), 
the Galactic sample may not be complete. A simple mass ratio
argument, if the stellar populations of both Galaxies are similar,
would predict a ratio of 10\,-\,12 \cite{sun92}
more X-ray binaries in the Milky
Way than in the LMC. Therefore the low number of
wind-fed systems in the LMC is not surprising in principle.

What sets out a difference is the fact that for one or two wind-fed
supergiants, there are 3 RLO persistent bright sources
in the LMC. This is in contrast to the Galaxy, where only one such
system (two RLO sources with massive companions, if the transient
black hole candidate \object{SAX J1819.3$-$2525} is counted) is
known. Moreover, the three donors in
the LMC RLO sources are quite close to the main sequence, while the
donor in \object{Cen X-3}, as the donor in \object{SMC X-1} -- the only
similar source in the Small Magellanic Cloud --, is rather more
evolved. Since no selection effects are expected to be so strong as to
affect the number of persistent sources with $L_{\rm
  x}\approx10^{38}\:{\rm erg}\,{\rm s}^{-1}$ detected in our Galaxy,
and a mass ratio argument would predict a much larger number of RLO
objects, there is a strong suggestion here of some significant
difference. Though the numbers involved are too small to attempt any
statistics, there is good reason to suspect that, for some reason,
binary evolution is more likely to result in close binaries with black
holes in the LMC than in the Milky Way, as has been discussed by
several authors (e.g., Johnston et al. 1979; Pakull 1898). 
This seems to be the only significant difference among the two
populations and it is unclear whether it can be assigned to the effects
of lower metallicity in binary evolution (e.g., through weaker stellar
winds resulting in higher pre-supernova core masses) -- see Helfand \&
Moran (2001) for a more thorough discussion.

 Among the objects in our LMC sample, the fraction of BeXBs is 71\%. If
all the proposed HMXBs in Table~\ref{tab:unknowns} are added, the
proportion is 70\%. The corresponding fraction for the Galactic sample
(i.e. BeXB/HMXB) is 58\% for systems with optical counterparts and 65\%
for all systems (i.e., identified + candidates). The fact that the
fraction is so similar in both galaxies strongly suggests that the
selection effects dominating our knowledge of the HMXB population are
similar -- i.e., in both cases, the main bias results from the
incompleteness of the Be/X-ray binary sample, due to their transient
X-ray source condition.

\begin{table}[ht]
\caption{ Be/X-ray binaries in the LMC with their spectral types and
  other known parameters. See the text for references.}
\begin{center}
\begin{tabular}{lccc}
\hline
Name & $P_{{\rm s}}$(s) & Spectral & Max $L_\mathrm{x}$\\
& & Type &  (erg s$^{-1}$)\\
\hline
\object{CAL 9} & $-$ & B0V & $7\times10^{34}$\\
\object{CAL E} & 4.1 & B0V& $4\times10^{37}$\\
\object{RX J0520.5$-$6932} & $-$ & O9V & $5\times10^{34}$ \\
\object{RX J0529.8$-$6556} & 69.5 & B0.5V& $2\times10^{36}$ \\
\object{EXO 0531.1$-$6609} & 13.7 & B0.7V& $1\times10^{37}$ \\
\object{RX J0531.5$-$6518} & $-$ & B2V &$3\times10^{35}$ \\
\object{RX J0535.0$-$6700} & $-$ &B0V& $3\times10^{35}$ \\
\object{1A\,0535$-$66} & 0.07 & B0.5III& $1\times10^{39}$\\
\object{1SAX J0544.1$-$7100} & 96.1 & B0V & $2\times10^{36}$\\
\object{H0544$-$665} & $-$ & B0V & $1\times10^{37}$\\
\hline
\end{tabular}
\end{center}
\label{tab:bex}
\end{table}

\subsection{The spectral distribution of Be/X-ray binaries}

The spectral distribution  of the optical counterparts to Galactic
Be/X-ray binaries was studied by Negueruela (1998), who found it to be
very different from the spectral distribution of isolated Be stars. 
The spectral distribution of optical counterparts peaks sharply at B0
and does not extend beyond B2 (roughly corresponding to
$\approx10\,M_{\sun}$). Such distribution can be explained if one
assumes that during the mass transfer phase previous to the formation
of the Be/X-ray binary (see next Section) material lost from the
system carries away a large amount of angular momentum \cite{vbv97}.

\begin{figure}
\begin{picture}(250,360)
\put(0,0){\includegraphics{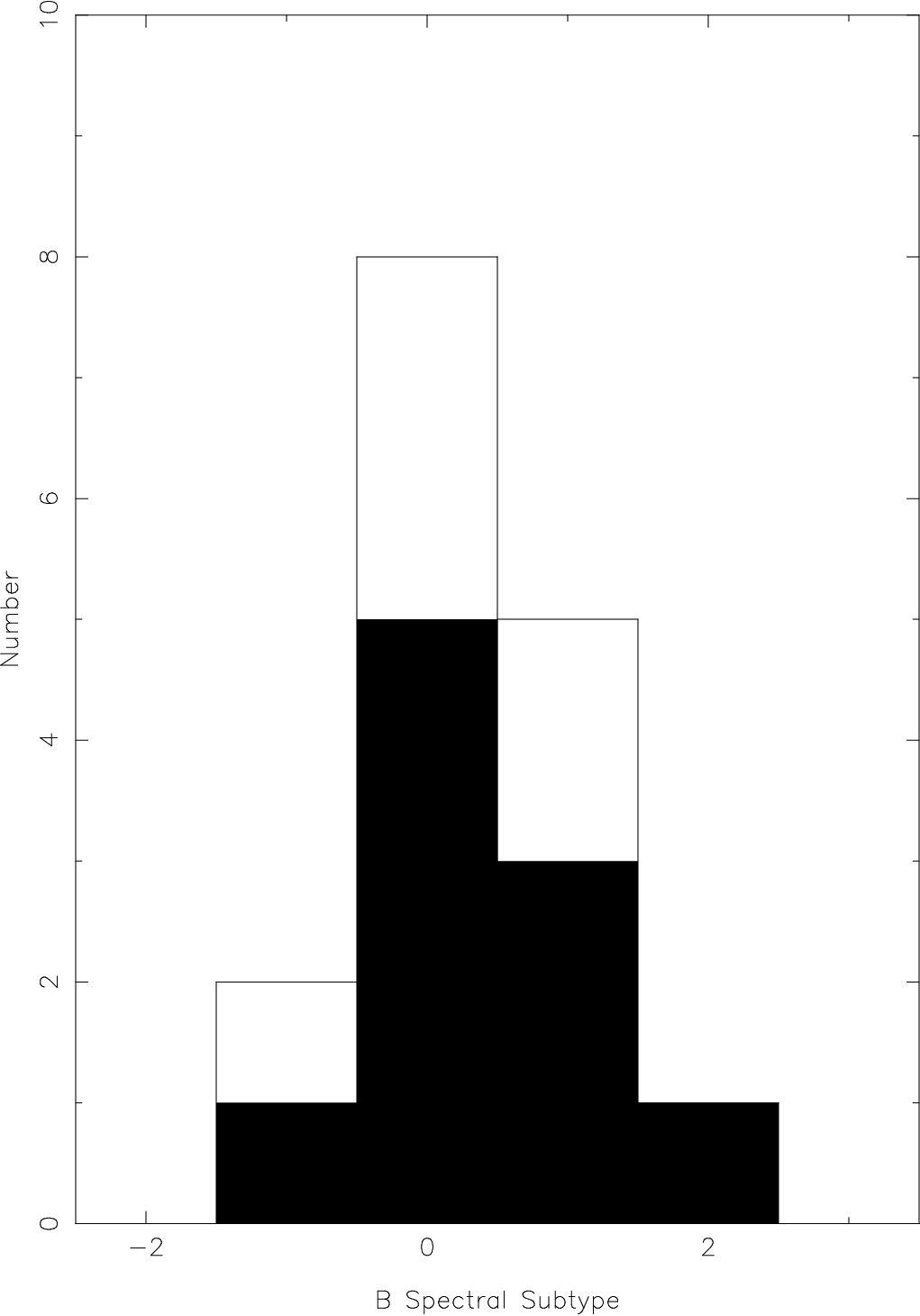}}
\end{picture}
\caption{The spectral distribution of the optical counterparts to
  Be/X-ray binaries in the LMC (filled) is compared to the
  distribution of 16 counterparts to Galactic Be/X-ray binaries
  identified with X-ray pulsars (hollow). Negative spectral subtypes
  are used to represent O-type stars.} 
\label{fig:hist}
\end{figure}

The spectral distribution of optical counterparts to LMC Be/X-ray
binaries is displayed in Fig.~\ref{fig:hist}, together with that of
Galactic sources. The convention adopted has been that of Steele et
al. (1998), i.e., the B0 bin contains objects with spectral types in
the range O9.5 to B0.2. In order to allow better comparison, the
Galactic sample has been replotted using the same convention. This
sample contains the 13 X-ray pulsars listed in Negueruela (1998) and
three new pulsars discovered by Motch et al. (1997): \object{RX
  J0440.9+4431}, \object{RX J0812.4$-$3114} and \object{RX
  J1037.5$-$5647}.  

A detailed statistical comparison is left for future work, after the
spectral types of some Galactic objects have been reassessed in view
of new high-quality spectra. However, it is clear from
Fig.~\ref{fig:hist} that the two distributions are basically
identical. Though such similarity is in principle expected (see Van
Bever \& Vanbeveren 1997), it is encouraging to see it confirmed,
because it must mean that:
\begin{itemize}
\item The Galactic sample, in spite of all the possible selection
  effects (see discussion in Negueruela 1998) is representative of the
  actual spectral distribution.
\item In spite of its small size (which could make us doubt of its
  statistical significance), the LMC sample represents well the
  population. 
\end{itemize}

If selection effects are having any impact on the spectral
distribution observed, such effects must affect both
samples. However, selection effects should be very different in
the Galactic plane (where we expect extinction towards the optical
companions to be the main source of bias in our knowledge of the
spectral distribution -- in the sense that earlier spectral types are
easier to observe and classify) and the LMC (where the main selection
effects are the brightness of the X-ray source and our ability to
actually identify the counterpart in crowded fields). Therefore we
conclude that the observed spectral distribution is not dominated by
selection effects. The only selection effect that could affect both
samples would be the existence of a population of objects with very
low X-ray luminosities, but such a population should be evident in the
Solar neighbourhood, if it was a major contributor to the numbers of
Be/X-ray binaries.

\subsection{Implications for binary evolution}

Massive X-ray binaries are born as the result of the evolution of a
massive binary in which mass transfer between the two components has
taken place before one of them (originally the most massive) became a
compact object after gravitational collapse of the core. In what
follows we will adopt the following convention: in the massive X-ray
binary, we will refer as the {\em donor} to the massive star which is
passing mass to the {\em compact object} (which will always be
referred to as such); in the original binary before the gravitational
collapse, we will call the {\em original primary} to the component
which evolved to become the compact object (and which was the more
massive component when the system reached the ZAMS) and the {\em
  original secondary} to the component that would later become the donor.

\begin{table}[ht]
\caption{Other suggested X-ray binaries in the LMC, not included in
  this sample. References are Cowley et al. (1997) = C97, Haberl \&
  Pietsch (1999) = HP99 and Sasaki et al. (2000) = SHP00.}
\begin{center}
\begin{tabular}{lcc}
\hline
Name & Suggested & Reference\\
& Counterpart & \\
\hline
\object{RX J0516.0$-$6916} & B1V & C97 \\
\object{RX J0512.6$-$6717} & $-$ & HP99\\
\object{RX J0532.4$-$6535} & {\object GRV\,0532$-$6536} & HP99\\
\object{RX J0535.8$-$6530} & $-$ & HP99\\
\object{RX J0541.4$-$6936} & \object{Sk $-69\degr$ 271} & SHP00\\
\object{RX J0541.6$-$6832} & \object{BI 267}& SHP00\\
\hline
\end{tabular}
\end{center}
\label{tab:unknowns}
\end{table}

Be/X-ray binaries are generally believed to be formed via
a standard evolutionary channel. The progenitor is an
intermediate-mass close binary with moderate mass ratio $q\ga 0.5$. 
The original primary starts transfering mass to its companion after
the end of the hydrogen core burning phase (case B), resulting in a
helium star and a rejuvenated main sequence star. 
If the helium star is massive enough, it 
will undergo a supernova explosion and become a neutron star. If the
binary is not disrupted, it can then become a Be/X-ray binary. This model 
has been developed by Habets (1987), Pols et al. (1991), 
Portegies Zwart (1995) and Van 
Bever \& Vanbeveren (1997), who have calculated the expected population 
distribution when different assumptions are made. All these models
assume that
the original primary must have a mass $\ga 12\, M_{\sun}$
(because otherwise it would not produce a neutron star) and that 
strong tidal interactions during the mass transfer  
phase result in the circularization of the
orbit. It is implicitly assumed that the Be nature of 
the original secondary is due to accretion of 
high-angular-momentum material from the primary, even though no
description of the exact physical process has been attempted. 

Habets (1987) studied possible evolutionary scenarios and came to the
conclusion that, if supernova explosions are always symmetric, only 
low-eccentricity Be/X-ray binaries may exist, because the exploding helium
star has in all cases a much lower mass than its companion and
therefore only a small
fraction of the system mass is lost in the explosion. As a
consequence, van den Heuvel \& van Paradijs (1997) conclude that 
the observational detection of Be/X-ray binaries with very eccentric
orbits is proof of the existence of intrinsic kicks imparted to the
neutron stars during the collapse that leads to their formation. 

A second evolutionary channel has been explored by Habets (1987). In
this case, the Be/X-ray binary is formed as the result of the evolution 
of a binary containing a massive star ($M_{{\rm MS}} \ga 20\,M_{\sun}$ where
$M_{{\rm MS}}$ is the main sequence mass of the primary) and
a rather less massive companion (of mass $M_{*} \sim
10\,M_{\sun}$). Because of the $q \la 0.5$ mass ratio, when the
primary fills its Roche lobe, mass transfer is highly
non-conservative, a common envelope forms and essentially all of the
hydrogen-rich envelope of the primary is lost from the system. The
resulting system consists of a relatively massive He star 
($M_{{\rm He}} \sim 5\,-\,10\: M_{\sun}$) and an unaffected
secondary. Since the helium star still contains a substantial fraction
of the total binary mass, a symmetric supernova explosion can now
result in a neutron star orbiting a Be star in a very eccentric orbit,
($e = M_{{\rm lost}}/M_{{\rm left}}$, where $M_{{\rm lost}}$ is the
mass lost from the system during the explosion and 
$M_{{\rm left}}= M_{*} + M_{{\rm x}}$ is the mass of the two
components after the explosion). 

An important limitation found by Habets (1987) is that, because of the
initial conditions required, this channel may only  
explain Be/X-ray binaries in which a neutron star orbits a low-mass Be
star in a close, eccentric orbit. This channel cannot produce systems with
either massive donors ($M_{*} \ga 15\, M_{\sun}$) or wide orbits. Also
important is the fact that, since the original secondary is left
basically unaffected by the whole process, it has to be implicitly
accepted that it becomes a Be star for some {\em intrinsic} reason,
unlike in the first channel considered.

The spectral distribution found for Be/X-ray binaries in both the
Milky Way and the LMC, sharply peaked at spectral type B0
(roughly corresponding to $M_{*}\approx 16\,M_{\sun}$) and not extending
beyond B2 ($M_{*}\approx 10\,M_{\sun}$) strongly argues against the
second channel contributing significantly to the formation of Be/X-ray
binaries. It is therefore strong indirect evidence in support of the
existence of supernova kicks.

In spite of this, the second channel is not only physically
possible, but almost certainly realized: a system like LMC X-3 (in
which the donor is {\em not} a Be star) 
must have formed through such a process, since the original primary must
have been sufficiently massive to result in a black hole and the
original secondary cannot have been more massive than the present 
$M_{*}\la 8\,M_{\sun}$. Therefore, in the original system, $q\la0.3$ and
non-conservative evolution must have occurred -- perhaps through case C
mass transfer, as in the models by Brown et al. (2001), since 
Wellstein \& Langer (1999) argue that cases A and B cannot result in
black holes.

Though it is not impossible that some of the Be/X-ray binaries with
lower-mass donors may have formed through the second channel, it is
clear that most must have followed the first channel. Taken at face
value, the dearth (or even absence) of Be/X-ray binaries forming
through the second
channel could be interpreted as a suggestion that a star does not
develop Be characteristics due to intrinsic reasons, but must have
undergone a process of mass transfer in a binary, as discussed by Gies
(2000). Such scenario is in disagreement with the population synthesis
models of Portegies Zwart (1995) and van Bever \& Vanbeveren (1997),
which indicate that not all Be stars may have formed through binary
evolution and therefore support an intrinsic cause for the Be
phenomenon. The dominance of the first channel may then simply be
a reflection of a preference for binary systems which originally have
a mass ratio $q\approx1$. If the progenitor binaries that will evolve
through the second channel are much less numerous than those which
follow the first, their end products will naturally be also a
minority. 

It is, however, intriguing to note that, while 
{\em no} system containing a Be star and a black hole is known,
several systems containing (non-emission) OB stars and a black hole
are known, and their donors span a wide range of spectral types. Since
their number is still relatively low, one can attribute the absence of
Be + black hole binaries to low number statistics. This is, in
principle, a valid argument, but it must be noted that black hole
binaries with a Be companion should have a much longer lifetime as an
X-ray source than any OB + black hole binary, because a mechanism for
mass transfer (the Be disk) exists during a longer period. The
situation should not be very different from binaries containing a
neutron star, where close to 70\% of systems known are Be/X-ray
binaries. Therefore the absence of known Be + black hole binaries
strongly suggests that there is some physical reason why Be + black
hole binaries cannot be formed or are not bright X-ray sources.

\begin{acknowledgements}

We acknowledge with great pleasure the help of
Nick Haigh, Jos\'{e} Miguel Torrej\'{o}n and Silas Laycock with some of the
observations. We also thank Drs J.~Simon~Clark and Manfred W. Pakull
for many useful comments on the
manuscript.\newline
This paper uses observations made from the South African Astronomical
Observatory (SAAO). This research has made use of the 
Simbad data base, operated at CDS,
Strasbourg, France. 
During part of this work IN was supported by an ESA
external fellowship.

\end{acknowledgements}

\end{document}